\documentclass[a4paper,fleqn,usenatbib]{mnras}

\usepackage{newtxtext,newtxmath}

\usepackage[T1]{fontenc}
\usepackage{ae,aecompl}

\usepackage{graphicx}	
\usepackage{amsmath}	
\usepackage{amssymb}	

\title[A Submillimeter Galaxy Projected on the Debris Disk of HD 95086]{A Submillimeter Background Galaxy Projected 
        on the Debris Disk of HD95086 Revealed by ALMA}

\author[Zapata et al.]{
Luis A. Zapata$^{1}$\thanks{E-mail: l.zapata@irya.unam.mx},
Paul T. P. Ho$^{2,3}$, and
Luis F. Rodr\'\i guez$^{1}$
\\
$^{1}$Instituto de Radioastronom\'\i a y Astrof\'\i sica, UNAM, Apdo. Postal 3-72 (Xangari), 58089 Morelia, Michoac\'an, M\'exico \\
$^{2}$Academia Sinica Institute of Astronomy and Astrophysics, PO Box 23-141, Taipei, 10617, Taiwan\\
$^{3}$East Asian Observatory, 666 N. A'ohoku Place, Hilo, Hawaii 96720, USA}

\date{Accepted XXX. Received YYY; in original form ZZZ}

\pubyear{2015}

\begin{document}
\label{firstpage}
\pagerange{\pageref{firstpage}--\pageref{lastpage}}
\maketitle

\begin{abstract}
We present sensitive observations carried out with the Atacama Large Millimeter/Submillimeter Array (ALMA) of the dusty debris disk HD 95086.
These observations were made in bands 6 (223 GHz) and 7 (338 GHz) with an angular resolution of about 1$''$ which allowed us to resolve well the 
debris disk with a deconvolved size of 7.0$''$ $\times$ 6.0$''$ and with an inner depression of about 2$''$. We do not
detect emission from the star itself and the possible inner dusty belt.   We also do not detect CO (J=2-1) and (J=3-2) emission, 
excluding the possibility of an evolved gaseous primordial disk as noted in previous studies of HD95086.  We estimated a lower limit for the gas mass of $\leq$0.01 M$_\oplus$
for the debris disk of HD95086. From the mm. emission, we computed a dust mass for the debris disk HD95086 of 0.5$\pm$0.2 M$_\oplus$, resulting in a dust-to-gas ratio of $\geq$50. 
Finally, we confirm the detection of a strong submillimeter source to the northwest of the disk (ALMA-SMM1) revealed by recent ALMA observations. 
This new source might be interpreted as a planet in formation on the periphery of the debris 
disk HD 95086 or as a strong impact between dwarf planets. However, given the absence of the proper motions of ALMA-SMM1 similar to those 
reported in the debris disk (estimated from these new ALMA observations) and for the optical star, this is more likely to be a submillimeter background galaxy.  

\end{abstract}

\begin{keywords}
Methods: observational -- 
Techniques: interferometric --\\ 
Protoplanetary disks -- HD95086\\
\end{keywords}



\section{Introduction}

The dusty debris disks around main sequence stars are thought to be produced from repeated impacts of large
planetesimals, as massive as the dwarf planets, asteroids or comets observed in our own Solar System \citep{mat2014}.  
These repeated impacts might translate in sudden changes of brightness ({\it e.g.} at infrared or submillimeter wavelengths) 
in some portions of the debris disks, see for example \citet{tele2005}. 

For younger stars, debris disks may contain primordial material, but at this moment this possibility is still not clear \citep{wya2008}.
It is thought that the optically thin debris disks are shaped by large bodies possibly through collisions 
and gravitational perturbations \citep{mac2017}.  Hence, the physical properties of the debris disks provide 
essential information on their nature and the circumstellar disk evolution \citep{mat2014}. 

HD 95086 is a main-sequence \citep[class III, with an age between 10$-$17 Myr and early-type A8 dusty star;][]{ram2013}.
{HD 95086 is located at a distance of 83.8 $\pm$ 1.9 pc \citep{gaia2016}, which is on the near side of the Lower Centaurus Crux
association \citep{dez1999}.  We have adopted this distance throughout the paper. }   
  
Very recently infrared (L$'$ band) observations made by \citet{ram2013} reported the discovery of a probable 4$-$5  M$_{Jup}$ 
giant planet around HD 95086, the exoplanet with the lowest mass ever imaged around a star. This exoplanet is located to the southeast
of the star at a projected distance of about 50 AU.   

HD 95086 harbours a massive debris disk (0.5 M$_\oplus$), first traced by a large infrared excess \citep{riz2012,che2012}
then resolved with Herschel/PACS observations \citep{moo2013}. APEX/LABOCA observations also 
detected this debris disk at submillimeter wavelengths \citep{nil2010}, but could not resolve it.  Modeling the spectral energy distribution and images of the system 
suggests a three-component debris structure \citep{su2015}. The exoplanet (HD 95086b) may be responsible for sculpting the outer edge of the debris gap 
\citep{ram2013,su2015}. This discovery added another system to the relatively small population of young, nearby, 
intermediate-mass stars hosting both debris disks and directly imaged giant planets \citep{su2015}.

\begin{figure*}
	\includegraphics[scale=0.33]{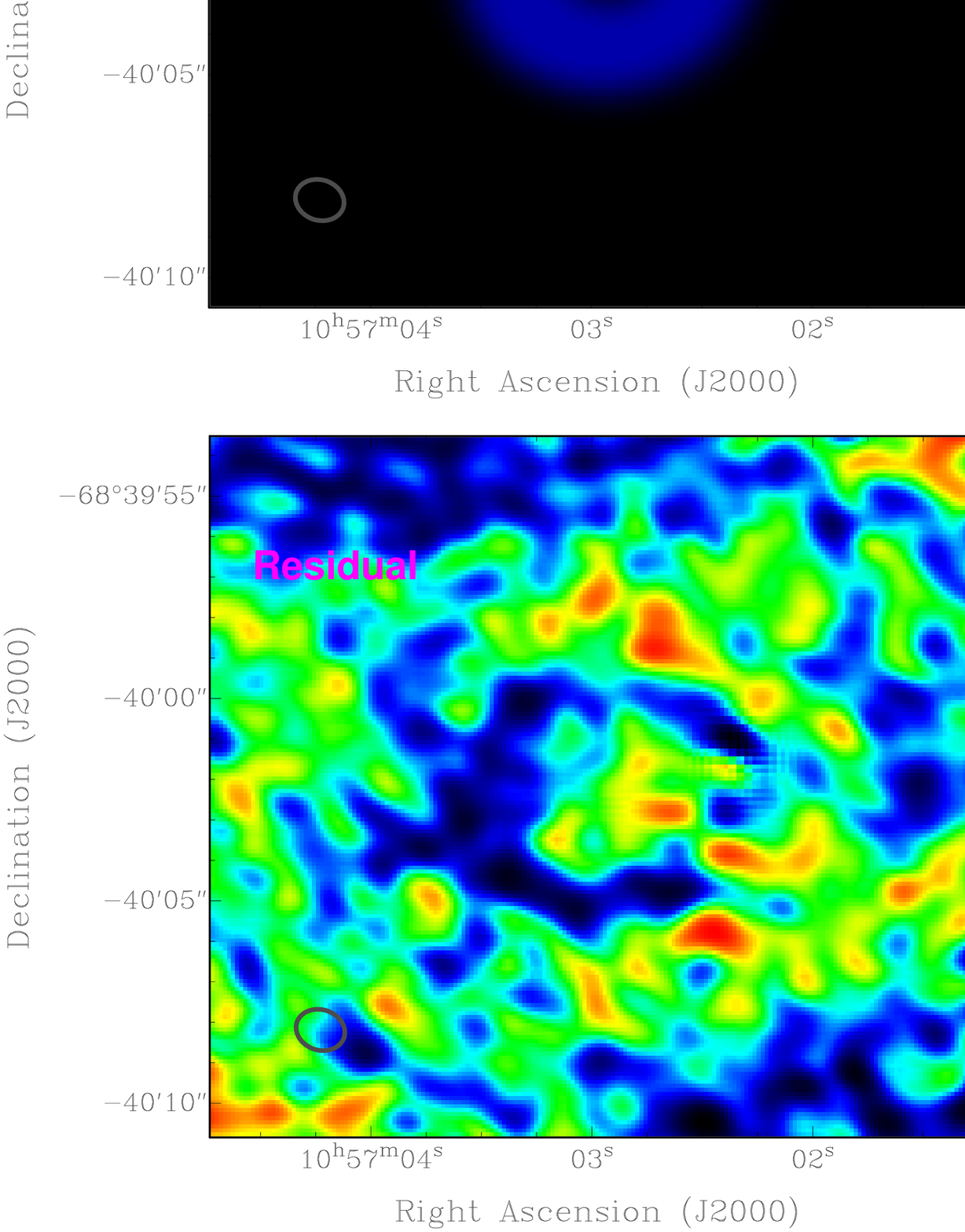}
	\includegraphics[scale=0.33]{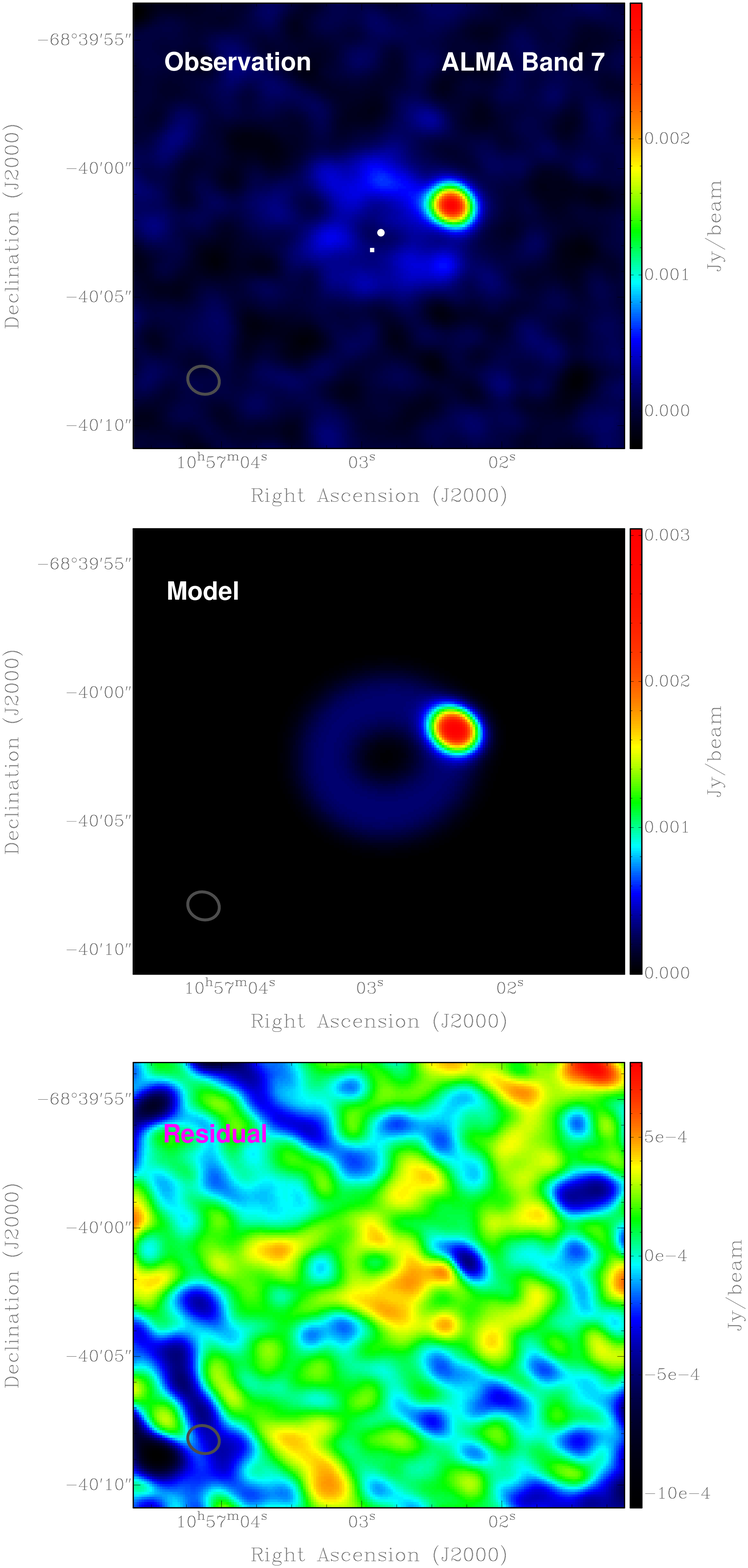}
         \caption{{Images of the ALMA band 6 and 7 observations (upper), synthetic models (middle), and residuals (lower) of the debris disk in HD 95086.  
                       The synthesized beam size is shown in the bottom left corner of each panel. The peak intensity scale-bar is shown in each panel.
                       The white filled circle and square mark the position of the star and exoplanet (HD 95086 b), respectively, reported in \citet{ram2013}. The positions of the star 
                       and the exoplanet are corrected for proper motions.}}
        \label{fig:Figure1}
\end{figure*}

\begin{table*}
	\centering
	\caption{Parameters of the debris disk and the background galaxy from our synthetic model}		
	\begin{tabular}{ccccccc} 
		\hline
		\hline
		 Observation          & Frequency & $\alpha_{2000}$ &  $\delta_{2000}$                                        &  Density Flux &  Deconvolved Size & Position Angle \\
		  Date                     &    (GHz)        &  (h m s)               &  ($^\circ$ $^{\prime}$  $^{\prime\prime}$) &  (mJy) &  ($^{\prime\prime}$) & ($^\circ$)\\ 
		\hline \\
		  & & & HD 95086 & & & \\ 
		  2015/01/28 (2015.08) & 223 & 10 57 02.920$\pm$0.05 & $-$68 40 02.37$\pm$0.35  & 6.0$\pm$0.3 & 7.0$\pm$0.6 $\times$ 6.0$\pm$0.4 & 90$\pm$10 \\
		  2017/04/26 (2017.32) & 338 & 10 57 02.880$\pm$0.04 & $-$68 40 02.44$\pm$0.29  & 21.8$\pm$1.1 & 7.0$\pm$0.9 $\times$ 5.8$\pm$0.8 &  95$\pm$15 \\
		\hline\\	
		& & & Background Galaxy & & & \\ 
		 2015/01/28 (2015.08) & 223 & 10 57 02.345$\pm$0.002   & $-$68 40 01.41$\pm$0.03  &  0.97$\pm$0.02 & 0.6$\pm$0.1 $\times$ 0.3$\pm$0.1 &  19$\pm$10 \\ 
		 2017/04/26 (2017.32) & 338 & 10 57 02.351$\pm$0.003   & $-$68 40 01.42$\pm$0.03  &  3.4$\pm$0.2   & 0.6$\pm$0.2 $\times$ 0.5$\pm$0.2 &  93$\pm$15 \\
		\hline
		\hline
		\multicolumn{7}{l}{For the background galaxy, the accuracy of the position (0.03$"$)  is a factor of two smaller than the ALMA theoretical astrometric accuracy of 0.06$''$.  See the} \\ 
		\multicolumn{7}{l}{ {ALMA Technical Handbook, section 10.6.3.\footnote{https://almascience.nrao.edu/documents-and-tools/cycle4/alma-technical-handbook.}}}
	\end{tabular}
	  \label{tab:Table1}
\end{table*}

Very recently, \citet{su2017} reported the detection of a bright source located near the NW edge of the debris disk. They discussed four possibilities
for the nature of the bright and compact source: a new phase of a circumplanetary disk, a dusty clump due to a giant impact, a dust clump due to 
planetesimals trapped by an unseen planet, and the alignment of a background galaxy. After a discussion of these possibilities, they arrived to the possibility that given   
its brightness at 1.3 mm and the potential spectral energy distribution are consistent with it being a luminous star-forming galaxy at high redshift.
However, it is not conclusive from their data.

We present ALMA band 6 and 7 sensitive high angular resolution ($\sim$1.0$''$) observations 
of the debris disk HD95086.  The disk is well-resolved with a physical size of about 7.0$''$ $\times$ 6.0$''$ and with an inner
depression of about 2.0$''$ .  {The observations also revealed the strong submillimeter source to the northwest 
of the disk, which we propose a submillimeter background galaxy.}

\section{ALMA Observations}

\subsection{1.3~mm (Archive)}

The ALMA\footnote{ALMA is a partnership of ESO (representing its member states), 
NSF (USA) and NINS (Japan), together with NRC (Canada) and NSC and ASIAA (Taiwan) 
and KASI (Republic of Korea), in cooperation with the Republic of Chile. 
The Joint ALMA Observatory is operated by ESO, AUI/NRAO and NAOJ.}  Band 6 (1.3 mm) observations were 
carried out between 2015 Jan 28 and April 6 as part of the Cycle 2 program 2013.1.00773.S. 
The observations used between 36 to 38  antennas with a diameter of 12 m, yielding baselines with projected lengths 
from 15 to 348.5~m (11--268~k$\lambda$).  The integration time on-source (HD 95086) was 4.84 hours.  
{We concatenated all the six epochs to obtain the best possible image.}

The phase center for all observations was the same $\alpha(J2000) = 10^h~ 57^m~ 02\rlap.^s907$;
$\delta(J2000) = -68^\circ~ 40'~ 02.66''$. The continuum image was obtained averaging line-free channels
from four spectral windows (of 1.875 and 2.000 GHz width) centered at rest frequencies: 230.005 GHz (spw0),  232.497 
GHz (spw1), 214.997 GHz (spw2), and 216.997 GHz (spw3), which covers a total bandwidth of 7.875 GHz. 
The spectral setup for spw 0 (at 230 GHz) contains the CO (2-1) line, but such line was not detected at a 4$\sigma$ level
of {1.0 mJy km s$^{-1}$}. {The spectral resolution for the spw 0 is 0.6 km s$^{-1}$.} 

The weather conditions were very good and stable with an average precipitable water vapor of about 1.2 mm 
and an average system temperature of 80 K. The ALMA calibration included simultaneous observations of the 
183 GHz water line with water vapor radiometers, used to reduce the atmospheric phase noise. 
Quasars J1107$-$4449, J1107$-$448, and J1145$-$6954 were used to calibrate the bandpass, the flux scale, the atmosphere 
and the gain fluctuations. 

The data were calibrated, imaged, and analyzed using the Common Astronomy Software Applications  \citep[CASA:][]{mac2007}.
Imaging of the calibrated visibilities was done using the task CLEAN. We combined all epochs with the task 
UVCONCAT in CASA. The resulting image rms noise was 10 $\mu$Jy beam$^{-1}$ at an angular 
resolution of $1\rlap.{''}2 \times 1\rlap.{''}0$ with a PA = $+$72.3$^\circ$.  The ALMA theoretical rms noise for this configuration, 
integration time, and frequency is 8 $\mu$Jy beam$^{-1}$, which is very close to the value we obtain in the continuum images.
 We used the ROBUST ({ in the Briggs weighting}) parameter of CLEAN in CASA set to 0
for an optimal compromise between angular resolution and sensitivity. The millimeter continuum image obtained 
for HD95086 was corrected by the primary beam attenuation. The primary beam at this wavelength had a full width 
at half maximum of about $27^{\prime\prime}$.  The resulting 1.3 mm continuum image is presented in Figure ~\ref{fig:Figure1}.
We have tried to do self-calibration using the resulting mm image as model, but as the millimeter sources are very faint,  we did not obtain
good solutions. 

\subsection{0.8~mm (New observations)}

The ALMA Band 7 (0.8 mm) observations were carried on 2017 Apr 26 as part of the Cycle 4 program 2016.A.00021.T. 
The observations used 41 antennas with a diameter of 12 m, yielding baselines with projected lengths 
from 16 to 460~m (20--575~k$\lambda$).  The integration time on-source (HD 95086) was 6 min.
  
The phase center for these submillimeter observations was the same as the millimeter observations, see above.  
The continuum image was obtained averaging line-free channels
from four spectral windows (of 1.875 and 2.000 GHz width) centered at rest frequencies: 343.030 GHz (spw0),  333.851 
GHz (spw1), 332.018 GHz (spw2), and 345.848 GHz (spw3), which covers a total bandwidth of 7.875 GHz. 
The spectral setup for spw3 (at 345 GHz) contains the CO (3-2) line, but such line was also not detected at a 4$\sigma$ level
of {25 mJy km s$^{-1}$}. {The spectral resolution for the spw 3 is 0.4 km s$^{-1}$.} 

The weather conditions, as in the millimeter observations, were very good and stable with an average precipitable water vapor of about 0.4 mm 
and an average system temperature of 130 K. The ALMA calibration included simultaneous observations of the 
183 GHz water line with water vapor radiometers, used to reduce the atmospheric phase noise. 
Quasars J1107$-$4449, J1427$-$4206, J1136$-$6827, J1107$-$448, and J1145$-$6954 were used to calibrate the bandpass, pointing, 
atmosphere, the flux scale, and the gain fluctuations. 
As in the millimeter wavelengths, the data were calibrated, imaged, and analyzed using CASA.
The resulting image rms noise was 80 $\mu$Jy beam$^{-1}$ at an angular 
resolution of $1\rlap.{''}2 \times 1\rlap.{''}0$ with PA = $+$69.5$^\circ$. The ALMA theoretical rms noise for this configuration, 
integration time, and frequency is 70 $\mu$Jy beam$^{-1}$, which is very close to the value we obtain in the continuum images.
We used the UVTAPER parameter of CLEAN in CASA set to {\it outertaper}$=$$1\rlap.{''}0$ to obtain a better sensitivity 
losing some angular resolution.  The resulting 0.8~mm continuum image is presented in Figure ~\ref{fig:Figure1}.

\begin{figure}\centering
	\includegraphics[scale=0.34,angle=-90]{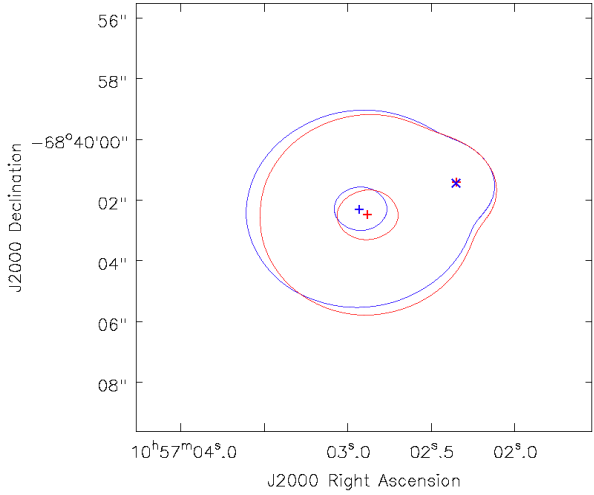}
         \caption{ {Contour images of the synthetic disk+galaxy models at bands 6 (blue) and 7 (red) of the debris disk HD 95086 made for the different epochs.  
                       The contours levels are 0.13 mJy (red) and 59 (blue) $\mu$Jy. 
                       The blue and red crosses mark the position of the phase center of the model at band 6 and 7, respectively.  The red and blue crosses to the northwest mark 
                       the positions of the submillimeter background galaxy at different epochs, see Table 1.  }}
        \label{fig:Figure2}
\end{figure}

\section{Results}

\begin{figure}\centering
	\includegraphics[scale=0.32,angle=-90]{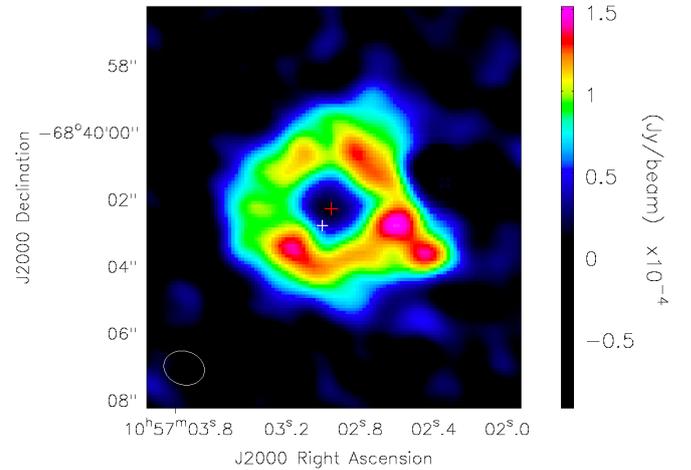}
         \caption{ ALMA continuum band 6 image of the debris disk HD95086 with the submillimeter background galaxy subtracted. The positions of the 
                       star and exoplanet (HD 95086 b) are marked with a red and white cross, respectively.  The peak intensity scale-bar is shown in the right. 
                        The synthesized beam size is shown in the bottom left corner.}
        \label{fig:Figure3}
\end{figure}

The main results of this study are presented in Figures ~\ref{fig:Figure1}, \ref{fig:Figure2}, \ref{fig:Figure3} and Table \ref{tab:Table1}. 

In Figure ~\ref{fig:Figure1}, we present the resulting images from the ALMA bands 6 and 7 high sensitivity observations
of HD 95086 (top panels).  From these top images one can see the well-resolved debris disk surrounding HD95086 already reported at
infrared and submillimeter wavelengths \citep{moo2013,nil2010}. Our images show that the debris disk is approximately rounded with a strong 
submillimeter compact source to its NW (ALMA-SMM1) and a deep inner depression. It has been estimated that this debris disk has an inclination of about 20$^\circ$ with respect to the 
plane of the sky  \citep{su2015}.  This explains its rounded morphology.  From our modeling, we found that the debris disk has 
similar dimensions to those estimated by \citet{moo2013} in the infrared regime ($\sim$ 6.0$''$ $\times$ 5.4$''$; 540 $\times$ 490 AU). We obtained a 
size for the disk of 7.0$''$ $\times$ 6.0$''$; 580 AU $\times$ 498 AU in band 6 and 7, see Table \ref{tab:Table1}. 
{The small difference between the two measurements (infrared and (sub)millimeter) can 
be explained proposing that at submillimeter wavelengths we are tracing colder dust located in the outermost parts within the debris disk. 
For the inner cavity of the disk, we estimated a size of 2.1$''$ $\pm$ 0.5$''$ $\times$ 1.7$''$ $\pm$ 0.4$''$ (174 AU $\times$ 140 AU), 
similar to the Herschel infrared observations \citep{su2015}.}

{In Figure ~\ref{fig:Figure1}, middle panels, we also include a synthetic model computed in CASA with the tasks SIMOBSERVE and SIMANALYZE.
This synthetic model was first computed in MIRIAD with the task IMGEN, where we generated a ring and a gaussian source at the position, and with dimensions of the debris disk
and the compact source, respectively. After this, we then iterated subtracting the model from the observations until we obtained noise in the residuals. { Here, we iterated using a trial and error
method, a more complete modeling is beyond the scope of this work.}  The best values for the ring and the compact source are presented in Table \ref{tab:Table1}. }      

Here, we do not need to know the value of the inclination angle of the ring, but from the values of the mayor and minor axis, it 
should be close to 20$^\circ$. Taking this synthetic image as a model, we then introduce it to SIMOBSERVE and SIMANALYZE in CASA to simulate 
the observations.  

{The dimensions of the modeled ring are $\sim$ 7.0$''$ $\times$ 6.0$''$ for the outer edge and 2.1$''$ $\times$ 1.7$''$ for the inner one. 
The middle panels show the modeled ring plus a Gaussian source obtained with CASA at different wavelengths. We used the 
same antenna configuration ({Cycle 2 for band 6 and Cycle 4 for band 7}) for the simulated observations.  In the lower panels of Figure ~\ref{fig:Figure1}, we show the residual images 
after subtracting the synthetic model of the debris disk HD95086 and the background galaxy, and the images obtained with ALMA. 
From this modeling, we obtained their physical parameters, see Table \ref{tab:Table1}. }  {We note that we have not used a background level term in the model.}  

In Figure ~\ref{fig:Figure2}, we present contour images from the modeled debris disk HD95086 at different bands and epochs. From this image one 
can note the displacement of the disk to its SW in the more recent observation at 345 GHz or band 7, see Table \ref{tab:Table1}. This displacement
is more evident in the disk centers marked with a blue and red cross in Figure ~\ref{fig:Figure2}.  {The positions of the disk centers from our models 
are given in Table \ref{tab:Table1}. From these positions and taking into account that the time difference between the observations is 2.24 years, we derived proper motions
of $\mu_\alpha$  $=$ $-$90$\pm$ 25 mas yr$^{-1}$ and  $\mu_\delta$  $=$ $-$31$\pm$ 27 mas yr$^{-1}$ for the debris disk.  These values are consistent to those proper motions reported by \citet{gaia2016} for the star, $\mu_\alpha$  $=$ $-$41.11$\pm$0.042 mas yr$^{-1}$ and  
$\mu_\delta$  $=$ $+$12.91$\pm$0.036 mas yr$^{-1}$.} Our large positional errors reflect the uncertainty of estimating the position of
an extended object (the debris disk) in the sky. However, for  ALMA-SMM1, the positional errors and its position in the sky are much better
determined.  From Table \ref{tab:Table1} and Figure ~\ref{fig:Figure2}, one can see that these positions are almost identical, suggesting that this source 
is a background submillimeter emitting source. ALMA-SMM1 is angularly unresolved in our two ALMA observations.

In Figure ~\ref{fig:Figure3} is presented an ALMA band 6 image of the debris disk surrounding HD95086, but subtracting in the {\it (u,v)} plane a point
source with the flux and position of ALMA-SMM1, see Table \ref{tab:Table1}. From this image it is better observed the
morphology of the debris disk, and its real mass can be better estimated. With the flux values presented in Table \ref{tab:Table1} for the debris disk, 
one can estimate a lower limit for the mass of the disk if the emission is optically thick.  Assuming that the dust emission is optically thin and isothermal 
at band 6, the dust mass (M$_d$) will be directly proportional to the flux density (S$_\nu$) as:

$$
M_d=\frac{d^2 S_\nu}{\kappa_\nu B_\nu(T_d)},
$$    

\noindent
where the $d$ is the distance to the object, $\kappa_\nu$ the dust
mass opacity, and B$_\nu(T_d)$ the Planck function for the dust
temperature T$_d$.  Assuming a dust mass opacity ($\kappa_\nu$) of 1.5
cm$^2$ g$^{-1}$, obtained for these wavelengths by \citet{Oss1994} 
for coagulated dust particles with no ice mantles. 
Assuming also an opacity power-law index 
$\beta$ = 1.0 (or $\alpha$ = 3.0 as estimated from our band 6 and 7 observations),
as well as a characteristic dust temperature (T$_d$) of 55 K \citep[for the cold component, see ][]{su2015}, we
estimated a lower limit for the dust mass for the disk of 0.5$\pm$0.2 M$_\oplus$. 
Note that the level of uncertainty in the mass lower limit 
here is only proportional to the estimation of the flux density, this could be very large (a factor of two or three) 
given the uncertainty in the opacities. This value for the mass agrees very well to that estimated in recent infrared and submillimeter studies, 
\citet{su2015,nil2010}. 

 Assuming that the $^{12}$CO(2-1) line emission is optically thin and in Local Thermodynamic Equilibrium (LTE), we can estimate an upper limit 
 for the mass of the debris disk  using the following equation \citep[e.g.,][]{sco1986,palau2007}, which for the (2-1) transition is given by:

\small
$$
\left[\frac{M_{H_2}}{M_\odot}\right]=7.6 \times10^{-16}\,T_\mathrm{ex}\,e^{\frac{16.59}{T_\mathrm{ex}}}
\,X_\frac{H_2}{CO} 
\left[\frac{\int \mathrm{I_\nu dv}}{\mathrm{Jy\,km\,s}^{-1}}\right]
\left[\frac{\theta_\mathrm{maj}\,\theta_\mathrm{min}}{\mathrm{arcsec}^2}\right]
\left[\frac{D}{\mathrm{pc}}\right]^2, 
$$ 
\normalsize

\noindent where we used 2.8 as mean molecular weight, and X$_\frac{H_2}{CO}$  is the abundance ratio between the molecular hydrogen and the carbon monoxide \citep[$\sim$10$^4$, e.g.,][]{sco1986}, $T_\mathrm{ex}$ is in units of K and assumed to be 20~K, $\int \mathrm{I_\nu dv}$ is the average intensity integrated over velocity ({here we take a $\mathrm{dv}$= 5 km s$^{-1}$, which has been observed in some other debris disks, see \citet{kos2013}}), $\theta_\mathrm{maj}$ and $\theta_\mathrm{min}$ are the projected major and minor axes of the disk (see Table \ref{tab:Table1}), and $D$ is the distance to the source (83~pc). We then estimate an upper limit for the gas mass for the disk of  $\leq$ 0.010 M$_\oplus$.

Taking the values obtained for the dust and gas mass for the debris disk HD95086, we estimated a dust-to-gas ratio of $\geq$50. 
{Here, we note that as the gas in debris disks is expected to have a secondary origin, {\it i.e.} thought to be released from 
comets and icy grains \citep{matraca2017,kra2017}, its X$_\frac{H_2}{CO}$ abundance
is rather lower than 10$^4$, so this difficulties more a correct estimation for the gas mass of the debris disks.  } 

\section{Discussion}

Given the results found in the previous Section, we suggest that it is likely that the new compact source (ALMA-SMM1) revealed
in these ALMA observations is a submillimeter background galaxy.  
This conclusion is mainly due to the absence of proper motions as those estimated 
in the debris disk and the optical star.  This is a similar case to that reported in $\epsilon$ Eridani using millimeter wave observations carried out with the 
Large Millimeter Telescope Alfonso Serrano (LMT) by \citet{cha2016}. This study also revealed the  
identification of numerous unresolved sources which could correspond to background dusty star-forming galaxies using
the relatively large proper motions of the debris disk and the star. These background objects are clearly observed
in their Figure 3 together with the positions of the debris disk at different epochs.  The detection of multiple background objects is possibly
due to the large field of view of the LMT. A second case where sensitive ALMA observations also revealed background objects
is the debris disc 61 Vir \citep{mar2017}. However, these background objects in 61 Vir have VLA and Herschel counterparts,
which more strongly suggests that they could be background radio galaxies.  One can think that future sensitive ALMA observations 
to more debris disks will surely reveal more background objects in their fields.   

How probable is that a background source with 3.4 mJy at 338 GHz falls inside the box of 6$''$ $\times$ 6$''$
that includes the debris disk of HD95086? following \citet{gea2017} the number of background
sources with a flux density of 3.4 mJy or higher at 850 $\mu$m is about $10^3$ per square degree.
Then, the a priori probability that such a source falls inside a box of 6$''$ $\times$ 6$''$ is only 0.003.  
 
We also looked for a counterpart at optical or X-rays at the position of the ALMA-SMM1 {in the SIMBAD catalogues, but we obtained negative results}.
{To obtain an accurate position of the star HD 95086 for different epochs we used the
position and proper motions given by the \citet{gaia2016}.} Assuming that the mean epoch of the ALMA
observations is 2015.17 we obtain the position $\alpha(2000) = 10^h~ 57^m~ 02\rlap.{^s}906; \delta(2000) = -68^\circ~ 40'~ 02\rlap.{''}26$ 
for this epoch and equinox 2000.0.
This position falls very close to the centroid of the hole of the debris disk (see Figure ~\ref{fig:Figure1}). HD 95086 has a giant exoplanet that
following the tradition is denominated HD 95086b. Using the offset reported by \citet{ram2013} we obtain a position of
$\alpha(2000) = 10^h~ 57^m~ 02\rlap.{^s}960; \delta(2000) = -68^\circ~ 40'~ 02\rlap.{''}81$ for this exoplanet (see Figure ~\ref{fig:Figure1} and  Table ~\ref{tab:Table1}).

In our search, we found an optical star very close to the new mm object, evident in the results of \citet{kou2005} and \citet{ram2013}.
{  It is obvious to note that the separation ($\sim$ 2.5 arcsecond) implies that these sources (the optical star and ALMA-SMM1) 
are not be related to each other.} 
In these papers the position of the star is given in terms of an offset from the position of HD 95086. From the
 \citet{kou2005} data, taken at several epochs with a mean value of 2000.93, we obtain a position of
$\alpha(2000) = 10^h~ 57^m~ 02\rlap.{^s}407; \delta(2000) = -68^\circ~ 39'~ 58\rlap.{''}90$
while from the \citet{ram2013} data taken at an epoch 2012.03 (2012 January 11) we obtain a position of
$\alpha(2000) = 10^h~ 57^m~ 02\rlap.{^s}384; \delta(2000) = -68^\circ~ 39'~ 58\rlap.{''}87$. In both cases the equinox is year 2000.0.
These two position differ by only $\sim0\rlap.{''}13$. If the background star had the same proper motions of HD 95086
we expected a displacement of $\sim0\rlap.{''}48$. We then conclude that the star is probably a more remote object than HD 95086.
{A similar conclusion was found already by \citet{kou2005}. }

\section{Conclusions}

We present sensitive ALMA observations of the debris disk around HD95086, a relatively young
star that harbours a directly imaged planet. Our conclusions about these observations are as follow:

\begin{itemize}

\item The high fidelity and sensitive ALMA maps revealed for the first time the structure of the debris disk around HD95086
         at submillimeter wavelengths. {From these images, we could estimate the disk with a size of 
         7.0$''$ $\times$ 6.0$''$ and with an inner depression of 2.1$"$ $\times$ 1.7$"$. These deconvolved sizes
         are similar to those reported at infrared and millimeter wavelengths.} 
          
\item  We detected no CO (J=2-1) and (J=3-2) emission, excluding the possibility of an evolved gaseous primordial disk. 
          We estimated a lower limit for the gas mass of $\leq$0.01 M$_\oplus$ for HD95086.
          From the submillimeter continuum emission, we estimated a dust mass for the debris disk of 0.5$\pm$0.2 M$_\oplus$,
          resulting in a dust-to-gas ratio of {$\geq$50}. 
          
\item  We confirm the detection of a strong submillimeter source to the northwest of the debris disk. Given the proper motions of the debris 
          disk as estimated from the new ALMA observations (and which are consistent to those reported for the optical star), we suggest that this 
          compact source is more likely to be a submillimeter background galaxy.             

\end{itemize}

\section*{Acknowledgements}
We are very thankful for the thoughtful suggestions of the anonymous referee that helped to improve our manuscript.
L.A.Z, and L. F. R. acknowledge the financial support from
DGAPA, UNAM, and CONACyT, M\'exico.  P.T.H  acknowledge the financial support from MOST 105-2112-M-001-025-MY3.
We would like to thank Yu-Ting Wu, who help us in preparing and reducing the ALMA band 7 observations.  
This paper makes use of the following ALMA data: ADS/JAO.ALMA\#2013.1.00773.S  and ADS/JAO.ALMA\#2016.A.00021.T.
This research has made use of the SIMBAD database, operated at CDS, Strasbourg, France \citep{wen2000}.









\bsp	
\label{lastpage}
\end{document}